\documentclass[12pt]{article}
\usepackage[margin=1in]{geometry}
\usepackage{times}
\usepackage{graphicx}
\usepackage{amsmath,amssymb}
\usepackage{cite}
\usepackage[labelfont=bf,hypcap=false]{caption}
\usepackage{authblk}
\usepackage{float}
\usepackage[hidelinks]{hyperref}

\usepackage{titlesec}
\titleformat{\section}{\normalfont\large\bfseries}{}{0pt}{}
\titleformat{\subsection}{\normalfont\normalsize\bfseries}{}{0pt}{}
\titlespacing{\section}{0pt}{12pt}{6pt}
\titlespacing{\subsection}{0pt}{8pt}{4pt}

\newif\ifshowfigures
\showfigurestrue
\newif\iffirstmainfigure
\firstmainfiguretrue
\newcommand{\mainfigure}[1]{%
  \ifshowfigures
    \iffirstmainfigure\global\firstmainfigurefalse\else\clearpage\fi
    \noindent\includegraphics[width=\textwidth]{submission_figures/#1}\par\nopagebreak
  \fi}
\newcommand{\figuresectiontitle}{\ifshowfigures Figures\else Figure legends\fi}

\title{\textbf{Floquet-driven light transport in programmable photonic processors via discretized evolution of synthetic magnetic fields}}

\author[1,*]{Andrea Cataldo}
\author[1]{Rohan Yadgirkar}
\author[1]{Ze-Sheng Xu}
\author[1]{Govind Krishna}
\author[2]{Ivan M. Khaymovich}
\author[1]{Val Zwiller}
\author[3,*]{Jun Gao}
\author[1,*]{Ali W. Elshaari}

\affil[1]{Department of Applied Physics, KTH Royal Institute of Technology, Albanova University Centre, Roslagstullsbacken 21, 106 91 Stockholm, Sweden}
\affil[2]{Univ. Toulouse, CNRS, Laboratoire de Physique Th{\'e}orique, Toulouse, France}
\affil[3]{School of Optical and Electronic Information, Huazhong University of Science and Technology, Wuhan 430074, People’s Republic of China}

\affil[*]{Corresponding authors. Emails: andreacl@kth.se, jungao@hust.edu.cn, elshaari@kth.se}

\date{} 

\begin{document}

\maketitle

\begin{abstract}
\noindent 
Photons, unlike electrons, do not couple directly to magnetic fields, yet synthetic gauge fields can impart magnetic-like responses and enable directional transport. Existing realizations rely on dedicated hardware, such as modulated resonators and strained lattices. Here, we present a protocol that realizes synthetic gauge fields on a general-purpose programmable photonic processor, and we validate it experimentally. A Floquet drive is discretized into ordered coupling steps with tunable hopping phases, and the time-ordered evolution is calculated and programmed on a universal interferometer mesh. On a 12-mode silicon processor, the measured dynamics exhibit chiral circulation that reverses under drive inversion with target populations above 99\%, flux-controlled routing, and directional flow stabilized by period optimization. These results quantify how accurately a general-purpose processor reproduces time-ordered, flux-dependent dynamics, and establish a programmable route to engineer gauge fields on available hardware.
\end{abstract}

\section{Introduction}

Magnetic fields profoundly alter the behavior of charged particles by reshaping their energy spectrum. In two-dimensional electron systems, a perpendicular field condenses continuous bands into a hierarchy of discrete Landau levels, each separated by the cyclotron frequency and hosting a macroscopic degeneracy of states. This quantization underlies the quantum Hall effect, wherein the electrical conductance assumes precisely quantized values. The effect also produces chiral edge states, which are one-way electron channels along material boundaries that are topologically protected against backscattering from disorder \cite{Thouless1982,Halperin1982}. Photons propagate through such magnetic fields unchanged, their spectra unquantized, their transport unprotected. Remarkably, synthetic magnetic fields can induce analogous phenomena in photonic systems, thereby enabling magnetic-like responses and robust optical transport  \cite{Onur2011,Fang2012_1,Hafezi2013}.

Several approaches have emerged to create these synthetic magnetic fields, broadly following two strategies: structural engineering and temporal modulation. Structural approaches exploit materials that break time-reversal symmetry or geometries that emulate gauge fields while preserving it. Magneto-optical materials provide non-reciprocal light propagation, wherein forward and backward traveling waves experience different phase shifts. Theoretical proposals \cite{Haldane2008,Raghu2008} and microwave demonstrations \cite{Wang2009} have validated this approach. At optical frequencies, however, magneto-optical effects become vanishingly weak. Strain-engineered geometries constitute an alternative approach. Carefully designed lattice deformations create spatially varying coupling between optical modes, generating synthetic magnetic fields without magnetic materials. Theoretical work has established that such strain produces photonic Landau levels \cite{Rechtsman2013,Guglielmon2021}. These predictions have been confirmed through observations in strained honeycomb lattices \cite{Jamadi2020} and silicon photonic crystals \cite{Barczyk2024,Barsukova2024}. Temporal approaches, in contrast, can break reciprocity through harmonic modulation of optical properties. Such modulation targets either the coupling between resonators \cite{Fang2012_2} or refractive index within resonators \cite{Tzuang2014}. Diverse platforms have realized this principle, from optomechanical systems \cite{Schmidt2015} to integrated electro-optic devices \cite{Dutt2020,Dinh2024}.

In this work, we present a protocol that realizes synthetic magnetic fields on a general-purpose programmable photonic processor, and we validate it experimentally. The protocol is implemented on a reconfigurable Mach-Zehnder interferometer (MZI) mesh, which supports arbitrary linear optical transformations \cite{Reck94,Clements16} and has enabled applications in quantum information processing \cite{Carolan15,Qiang2018,Paesani2019,Wang2020}, neuromorphic computing  \cite{Shen2017,Shastri2021}, and signal processing \cite{Perez2017, Marpaung2019}. We use a Floquet drive in which each period is divided into unitary substeps that sequentially activate selected pairs of lattice sites while other couplings are held off. The time-ordered product of these steps induces complex hopping phases for photons, realizing a synthetic magnetic flux. Because the sequence is non-palindromic and the substeps do not commute, the evolution depends on their order, which breaks time-reversal symmetry at the level of the programmed synthetic evolution. This does not imply optical nonreciprocity of the physical processor, which remains reciprocal during each programmed measurement. The ordered product is evaluated in software, and the mesh is programmed with the resulting evolution operator for each evolution time, as in earlier simulations on programmable processors \cite{Sparrow2018,Qiang2021,Wang2022}. Such programmability allows a single photonic processor to investigate various driven phenomena without separate fabrications by simply adjusting control signals \cite{Harris2017,Bogaerts2020}, building on recent implementations of programmable lattices in integrated photonic meshes \cite{On2024, Dai2024, Ma2024, Love2025}. Our processor belongs to the class of universal processors that are also available commercially \cite{Maring2024,Taballione2023}, and the protocol can be run on any of them.

The theory provides the drive protocol, the flux generated by the drive order and its dependence on the input site, the symmetry reduction of the seven-site lattice, and the selection of the drive period from the Floquet spectrum. For laser light, the measured populations follow from the calculated evolution operator. The experiment establishes how accurately a fabricated thermo-optic mesh, with imperfect couplers, loss and thermal crosstalk, reproduces these time-ordered and phase-sensitive dynamics. We implement three-, four-, and seven-site lattices. Our measurements reveal how each geometry produces distinct magnetic signatures, with the three-site system establishing the fundamental mechanism, the four-site system revealing flux interference and flux-controlled routing, and the seven-site system demonstrating directional photon transport in complex systems.

\section{Results}

\subsection{Chiral Floquet micromotion in a triangular plaquette}
Fig.~\ref{fig:Figure_1} shows a triangular plaquette, where each site represents one spatial waveguide mode of the processor. This geometry is the simplest closed loop in which the effect of the Floquet ordering can be seen directly. The system is initialized on site~1 and evolved over one modulation period \(T\). In this minimal case, each Floquet substep couples only one pair of modes, so the dynamics reduce to a sequence of two-mode Rabi transfers represented by the MZI sketches in Fig.~\ref{fig:Figure_1}a.

At \(t=0\), the programmed operation is the identity; all MZIs are set to full reflectivity, \(R=1\), so the population stays on site~1. During the first substep, the MZI implementing the \(1\leftrightarrow2\) coupling is tuned from \(R=1\) to \(R=0\). At \(t=T/3\), \(U_1(T/3)\) has been fully applied and the population has transferred from site~1 to site~2. The same idea is then applied to the next active bond; by \(t=2T/3\), \(U_2(T/3)\) transfers the population from site~2 to site~3. Finally, \(U_3(T/3)\) activates the \(3\leftrightarrow1\) coupling and returns the population to site~1 at \(t=T\).

The CW sequence in Fig.~\ref{fig:Figure_1}b thus drives coherent population transfer along \(1\rightarrow2\rightarrow3\rightarrow1\), while reversing the drive order produces the CCW path \(1\rightarrow3\rightarrow2\rightarrow1\) in Fig.~\ref{fig:Figure_1}c. The measurements closely follow the simulated dynamics and reach near-unity occupation at the expected switching times. This reversal is consistent with time-reversal-symmetry breaking within the programmed Floquet evolution.  

We also note that the observed circulation depends on the choice of initialization site. Starting from site 1 produces one-period CW/CCW cycles as demonstrated. If instead the system is prepared on site 2 or 3, the population moves only every other substep, so the effective circulation appears reversed and closes only after two periods (Supplementary Note~1.5). Thus, on average the effective magnetic field is zero, but it depends on both site location and the phase of the drive. Similarly, even for the same site~1 initialization, sparse stroboscopic sampling can introduce aliasing; under such sampling, the trajectory can appear reversed. We emphasize that this motion should not be interpreted as an input-independent magnetic response. 

\subsection{Flux-steered interferometry in a two-plaquette lattice}
The next step extends the single triangle to a pair of triangular plaquettes that share an edge, forming a two-path interferometer configuration (Fig.~\ref{fig:Figure_2}a). Both plaquettes are CW modulated with period \(T=3\pi/(2J)\). Peierls phases are programmed on the outer bonds to control the loop phases \(\Phi_A\) and \(\Phi_B\). The relevant quantity that steers interference at the shared vertices is the synthetic flux difference
\begin{equation}
\Delta\Phi = \Phi_A - \Phi_B,
\end{equation}
with the phase convention used in the implementation given in Supplementary Note~1.6. 

The input is prepared in the symmetric superposition \((|1\rangle + |3\rangle)/\sqrt{2}\) so that amplitudes launch into both loops. During one modulation period \(T\), the CW sequence routes these amplitudes to the shared sites \(2\) and \(4\) where they interfere. With this initial state, reversing the drive order yields the same population distribution at the shared sites, so the protocol's time-reversal symmetry breaking is obscured in this configuration. Using a different input, such as adding a relative phase between \(|1\rangle\) and \(|3\rangle\), would reveal the driving direction. For this reason, we use this configuration to probe the interferometric response rather than to demonstrate drive reversal.

The mid-period populations \(P_2(T/2)\) and \(P_4(T/2)\) are complementary and cross at \(\Delta \Phi=0\), where the populations are balanced. This follows from a two-path picture. During \(U_2\) the fields at the shared sites are coherent sums of the amplitudes from the two plaquettes that differ only by the programmed flux difference. With the phase origin chosen so that \(\Delta \Phi =0\) yields equal populations, the dependence is consistent with
\begin{equation}
P_{2,4}(T/2) = \frac{1}{2}\left[1 \pm V\sin(\Delta \Phi/2)\right],
\label{eq:mid_T}
\end{equation}
where \(0 \leq V \leq 1\) is the visibility, the plus sign corresponds to site 2, and the minus sign to site 4. This form explains the smooth S-shaped exchange of population.

Figures~\ref{fig:Figure_2}b--d show the full one-period dynamics for three representative settings of \(\Delta \Phi\). For \(\Delta \Phi = 0\) (CW/CW in the relative sense), the amplitudes from the two loops arrive in phase, distributing populations equally at sites 2 and 4 at the mid-period. Shifting to \(\Delta \Phi = + \pi\) (CW/CCW) reverses the relative sign of one path, so that constructive interference occurs at site 2 while destructive interference suppresses site 4. Reversing the sign to \(\Delta \Phi=-\pi\) flips the pattern, brightening site 4 and dimming site 2. The contrast is immediate in the measured traces.  

\subsection{Seven-site dynamics in the Rabi regime}
\label{sec:seven_site}
We now scale to a geometry where intertwined loops come together in a seven-site hexagon surrounding a central site (Fig.~\ref{fig:Figure_3}a). In the three-site and four-site lattices each step of the drive couples only two sites, so the evolution follows simple two-level dynamics with complete transfer via Rabi \(\pi\)-pulses. This clear condition no longer applies in lattices where a driving step may couple more than two sites, which breaks the isolated two-level picture and makes the effective Floquet Hamiltonian long-ranged.  

Nevertheless, the seven-site lattice has an inversion symmetry through the central site that can be used to recover a simpler Rabi regime. Under this symmetry, opposite perimeter sites are naturally described by symmetric and antisymmetric combinations; for example, the pair \((1,4)\) forms \((|1\rangle+|4\rangle)/\sqrt{2}\) and \((|1\rangle-|4\rangle)/\sqrt{2}\). The experiment uses the latter, for which coupling amplitudes from opposite sites into the center cancel. The dynamics are then confined to the antisymmetric subspace spanned by the sectors \(A=(1,4)\), \(B=(2,5)\), and \(C=(3,6)\). Details of the reduction are given in Supplementary Note~8.

At the Rabi \(\pi\)-pulse condition, the sectors are sequentially coupled. The grouped populations \(P_A=P_1+P_4\), \(P_B=P_2+P_5\), and \(P_C=P_3+P_6\) therefore follow \(A \to B \to C \to A\) for CW modulation, while reversing the substep order reverses the circulation. The measured traces closely follow this motion (Fig.~\ref{fig:Figure_3}b, c). 

This can be further quantified through the second-harmonic of the perimeter populations
\begin{equation}
C^{(2)}(t) = \sum_{k=1}^{6}P_k(t)e^{2i\theta_k},
\qquad
\phi^{(2)}(t)=\arg\!\big[C^{(2)}(t)\big]
\quad(\mathrm{mod}\ 2\pi),
\label{eq:phi_2}
\end{equation}
where \(\theta_k\) is the azimuthal angle of perimeter site \(k\). The second-harmonic phase \(\phi^{(2)}(t)\) provides a measure of the net angular drift and is expected to have an overall linear variation. We define \(\nu_2 \equiv \frac{1}{2\pi}\frac{d\phi^{(2)}}{d(t/T)}\) as the corresponding slope. From Fig.~\ref{fig:Figure_3}d we obtain \(\nu_2^{\text{CW}}=-0.504\) and \(\nu_2^{\text{CCW}}=+0.506\), with the sign reversing under drive inversion. Since the system is finite and open, \(\nu_2\) is not a quantized bulk winding number. The time-reversal symmetry breaking signature is the modulation reversal, not the choice of harmonic itself.

Because each evolution time is programmed as a single setting of the same mesh, the accuracy does not systematically degrade with the evolution time. The target populations are 99.58\%, 99.33\%, 99.21\% and 99.54\% after 0, 2, 4 and 6 substeps for CW modulation, and 99.48\%, 99.60\%, 99.52\% and 99.36\% after 0, 1, 3 and 5 substeps for CCW modulation (Fig.~\ref{fig:Figure_3}b,c).

\subsection{Seven-site dynamics in the period-optimized regime}
\label{sec:7S_single}
We now consider the case of a localized input \(|1\rangle\), where both antisymmetric and symmetric sectors contribute. The resulting evolution therefore contains both perimeter circulation and center-mediated oscillations, so the modulation period must be selected with both effects in mind. To do this, we identify stable operating points by examining the quasi-energy spectrum of the Floquet operator \(U_F\). Its eigenvalues take the form \(\lambda_k = e^{-i \varepsilon_k T}\), where the quasi-energies satisfy \(0 \le \varepsilon_k < 2\pi/T\). The stability of the dynamics is set by the separation between the quasi-energies, defined as the minimal splitting \(\Delta \varepsilon_\text{min} = \min_{i \neq j} |\varepsilon_i - \varepsilon_j|\), which quantifies the smallest gap in the spectrum (distance measured mod \(2\pi/T\) on a circle).

In a physical implementation, small coherent errors in control parameters (e.g. heater currents) can perturb the Floquet operator. A small gap (\(\Delta \varepsilon_\text{min} \rightarrow 0\)) indicates a near-degeneracy where the system's final state becomes more sensitive to such errors and Floquet modes can mix. A larger gap corresponds to a more stable, well-separated evolution, and provides a practical criterion for selecting the modulation period. Our strategy is therefore to find the optimal period that maximizes this gap
\begin{equation}
    T_\text{opt} = \arg \max_{T} \left [\Delta \varepsilon_\text{min}(T) \right].
\label{eq:T_opt}
\end{equation}

To validate the period-selection procedure, we first applied it to the three-site lattice. The results, shown in Fig.~\ref{fig:Figure_4}a, reveal a periodic spectrum of stability peaks. Notably, one peak \(\tau\) coincides with the familiar \(\pi\)-pulse condition, while the subsequent peaks represent its odd multiples, where a complete state transfer also occurs. This confirms our method correctly identifies the entire family of known coherent evolution points, each corresponding to a non-degenerate system with a minimal splitting of \(\Delta \varepsilon_\text{min} = 0.125\). 

We then evaluated the seven-site hexagon (Fig.~\ref{fig:Figure_4}b). Although more complex, the system retains features of the three-site lattice, including the shared near-degeneracy locations. An optimal peak \(JT_\text{opt} = 13.6647\) was selected for the experiment, which occurs near a \(3\tau\) stability peak of its underlying three-site building block. The slight shift from this value reflects the balance between two relevant coupling scales; a perimeter motion governed by \(x_\mathrm p = JT/3\), and a center-coupled motion governed by \(x_\mathrm c = \sqrt{2}JT/3\). At the selected period, \(x_\mathrm p\) is close to a \(3\pi/2\) Rabi transfer, while \(x_\mathrm c\) is close to a \(2\pi\) return. The optimized period therefore advances the perimeter population while bringing the center-coupled population nearly back by the substep boundary. An analytic period estimate is given in Supplementary Note~9.

Fig.~\ref{fig:Figure_4}(c--f) shows the resulting population dynamics over three modulation periods. Oscillatory exchange between coupled sites is superimposed on global directional flow. These oscillations migrate around the hexagon with each Hamiltonian switch. The stripe-like structure directly follows from the coexistence of perimeter motion and center motion. The CW case is shown in Fig.~\ref{fig:Figure_4}(c, d) and the CCW case in Fig.~\ref{fig:Figure_4}(e, f), where reversing the drive order flips the direction of the rotating pattern. Extended \(9T\) dynamics are provided in Supplementary Note~10.

This period-optimized regime retains nearly the same \(\pm 1/2\) second-harmonic phase drift observed in the Rabi regime. Thus, the Rabi and period-optimized regimes show distinct internal dynamics but nearly identical angular drift.

\section{Discussion}

We have presented a protocol that realizes synthetic gauge fields on a programmable photonic processor through discretized Floquet evolution, and validated it experimentally. Time-reversal symmetry is broken within the programmed synthetic evolution through non-palindromic, time-ordered coupling sequences. The physical picture comes from the theory, with the drive protocol, the flux generated by the drive order, and the selection of the drive period. For laser light the measured populations follow from the calculated evolution operator, and the experiment was not needed to predict them. It shows that a fabricated general-purpose mesh, despite imperfect couplers, loss and thermal crosstalk, faithfully reproduces the dynamics, with target populations above 99\% for both drive orders, an RMSE on the order of \(10^{-2}\), and no loss of accuracy as the evolution time increases. 

The processor used here belongs to the same class as other universal programmable processors, including commercial ones \cite{Taballione2021,Taballione2023,Maring2024}. Its mean fidelity for random 12-mode transformations is 98.1\% (Supplementary Note~6). This value is defined with the squared overlap of the measured and target intensity distributions; the fidelities reported for silicon nitride processors, 90.4\% for 12 modes \cite{Taballione2021} and 97.4\% for 20 modes \cite{Taballione2023}, use the overlap itself, which is always the larger of the two, and the highest reported value is 99.7\% for 12 modes \cite{Maring2024}. Each phase shifter needs 1.2~mW for a \(\pi\) shift \cite{Krishna2026}, against about 385~mW in the 12-mode silicon nitride processor \cite{Taballione2021}, so the mesh dissipates less than one watt and is stabilized with a thermoelectric cooler alone. The circuit, 66 MZIs on \(16\times3\)~mm\(^2\), is made in a silicon-on-insulator foundry process at telecom wavelength and is programmed with a model trained on intensity measurements with external detectors only. Silicon nitride has the lower propagation loss, 0.1~dB/cm against 2~dB/cm, which matters for experiments with many photons. The significance of the work lies in what it adds to such processors: synthetic gauge fields and Floquet drives become a function that is set in software, without modulators, magneto-optic materials or a lattice fabricated for the purpose. The lattice geometry, including couplings between non-adjacent waveguides, the loop phases, the drive order, the drive period, the input state and the evolution time are all programmed on one chip. Because each evolution time is programmed as a single transformation, the circuit depth, \(N(N-1)/2\) MZIs for \(N\) sites \cite{Clements16}, does not grow with the evolution time, in contrast to implementations in which every drive period is a physical section of the device \cite{Rechtsman2013Floquet,Maczewsky2017}.

The protocol requires only a universal multiport interferometer \cite{Reck94,Clements16}. It therefore applies to other processors and to larger lattices, up to twelve sites on the present chip. Several extensions may be considered without changing the protocol; multi-photon inputs, for which the same settings implement the dynamics of indistinguishable photons in the synthetic magnetic field \cite{Aaronson2013,Sparrow2018,Qiang2021}; loss through unused modes of the mesh \cite{Tischler2018,Cataldo2026}; and disorder set in software \cite{Harris2017}. One class of extensions is not covered; effects that act on the state between substeps, such as a Kerr nonlinearity, require physically successive substeps, e.g., in meshes with nonlinear elements between sections \cite{Shen2017,Bandyopadhyay2024}. 

\section{Methods}

\subsection{Theoretical framework}
Consider a programmable lattice governed by a periodic modulation cycle \(T\). We partition the evolution into a sequence of discrete unitary operations \(U_k\) (Fig.~\ref{fig:Figure_5}a,b), where each step activates a specific subset of couplings for a duration \(T/3\) (Fig.~\ref{fig:Figure_5}c; Supplementary Note~1.1). The product over one modulation period defines the Floquet operator. The clockwise (CW) and counter-clockwise (CCW) modulation orders are then simply the two ordered products, i.e. the same steps, reversed,
\begin{equation}
U_F^{\mathrm{CW}}(T)=U_3U_2U_1,
\qquad
U_F^{\mathrm{CCW}}(T)=U_1U_2U_3 .
\label{eq:UF_CW_CCW_main}
\end{equation}
Since the steps generally do not commute, i.e. \([U_k,U_{k'}] \neq 0\), reversing their order changes the Floquet operator. 

To capture the complete transient dynamics of this driven system, we define the total time-evolution operator \(U(t)\). For an arbitrary time \(t=nT+\delta t\), where \(n\) is the number of completed periods and \(\delta t\) is the remainder \(0 \leq \delta t < T\), the evolution is given by
\begin{equation}
    U(t)=U_\text{intra}(\delta t)[U_F(T)]^n,
\label{eq:Ut}
\end{equation}
where \([U_F(T)]^n\) describes the stroboscopic dynamics over full periods, and \(U_\text{intra}(\delta t)\) accounts for the residual intra-period evolution. 

For the stroboscopic dynamics, the same ordered product may be written in  terms of an effective Floquet Hamiltonian
\begin{equation}
H_F = \frac{i}{T}\operatorname{Log}\!\left[U_F(T)\right],
\end{equation}
after fixing the Floquet gauge, i.e. the logarithm branch and time frame (Supplementary Note~1.2). Complex hopping amplitudes in \(H_F\) can thus arise purely from the ordering of the constituent Hamiltonians. In the high-frequency limit, the leading order-dependent terms, when nonzero, are commutators between different substeps and reverse sign when the drive order is inverted (Supplementary Note~1.3). The ordered drive can therefore generate synthetic magnetic flux even without explicit Peierls phases; when present, these phases provide additional control, e.g. for tuning loop phases. We use the phase convention defined in Supplementary Note~1.4.

\subsection{Experimental setup}
The experimental platform (Fig.~\ref{fig:Figure_6}) was based on a 12-mode photonic processor. Here, the 12 modes represent independent spatial waveguide paths that constitute the Hilbert space of the system, where each waveguide acts as a discrete site in our programmed synthetic geometry. The processor integrates 264 elements, including thermo-optic phase shifters and multimode interferometers (MMIs), which together form the reconfigurable MZI mesh. Deep trenches were included around the MZIs to suppress thermal crosstalk between neighboring phase shifters. 

A continuous-wave laser provided the coherent light source at 1550 nm. Light was routed to individual chip inputs via a MEMS fiber optical switch optimized for TE polarization. Output signals were collected through a second identical switch connected to an optical power meter. A PC coordinated all instruments to provide full automation. Phase relationships were preserved by stabilizing the chip temperature using a thermoelectric cooler and PID controller. Detailed technical specifications of the experimental platform are provided in Supplementary Notes~2 and 3.

\subsection{Synthetic lattice realization}
The programmable photonic chip allowed us to implement three distinct lattice configurations. We first realized a three-site triangular lattice whose single closed loop allows the Floquet-induced circulation to be interpreted directly under CW and CCW modulation. We then programmed a four-site lattice consisting of two coupled triangular lattices with independent phase control. The shared vertices where pathways converge provided observation points for interference effects. Finally, we configured a seven-site hexagonal lattice to test synthetic magnetic fields in multi-loop geometries. Six perimeter sites surrounding a central node formed a network of interconnected triangular plaquettes, revealing whether directional transport persists in complex systems. 

Operating conditions varied across lattice configurations based on their specific requirements. For coherent population transfer in the three-site and four-site systems, we set coupling strengths and evolution times to achieve Rabi \(\pi\)-pulses. This operating point is robust to perturbations in a broad range, given by a finite gap in the corresponding spectrum of the Floquet operator, so the dynamics are tolerant to small period errors. The seven-site lattice was studied in two regimes, namely a Rabi regime and a period-optimized regime. The three-site and seven-site experiments tested both modulation sequences to establish chirality, while the four-site system used only CW driving to focus solely on the interference effects. 

\subsection{Unitary implementation}
We realized the evolution operators as target transformations on the programmable MZI mesh. Each MZI implements an SU(2) transformation, parameterized by two independent phase degrees of freedom. To map the theoretical models onto the device, we defined the \(n\) lattice sites as a subspace of the available spatial modes. The target \(n\times n\) operator was embedded into the processor's full Hilbert space by decoupling the inactive waveguides, ensuring the evolution remained confined to the synthetic lattice. 

The processor is based on a rectangular Clements mesh~\cite{Clements16}, which provides the programmable degrees of freedom needed to represent the transformations by a deterministic arrangement of local phase shifts. To ensure high-fidelity control, the programmed settings were found using a neural-network model of the processor, which accounts for hardware imperfections. Further details regarding the unitary implementation are found in Supplementary Notes~4--6.

\subsection{Data acquisition and analysis}
To resolve the transient population dynamics, we applied the unitary evolution operators incrementally. Each modulation period \(T\) was divided into 60 discretized steps (20 per substep). For a given time step \(t\), we computed the cumulative operator \(U(t)\). The corresponding current settings were found using a neural-network model of the photonic processor trained on random current settings and measured output powers, so that the predicted intensity response \(f(I)\) matched the target intensity matrix \(|U(t)|^2\). For experiments requiring a superposition input, the implemented unitary was composed of a preparatory unitary followed by the Floquet evolution. At each step, the programmed current settings were held fixed while output powers were measured across all active waveguides using an optical power meter. The transient dynamics were then reconstructed over repeated measurements of successive values of \(t\). For visual clarity, experimental markers in the line plots are shown every second sample; heatmaps and all reported target populations, fits, and RMSE values are computed using the full dataset. To account for fiber-to-chip coupling variations, the raw power readings were normalized to the total detected power at that time step. Uncertainties are reported as two-sided 95\% confidence intervals of the mean over repeated measurements, computed using Student's \(t\) distribution (see Supplementary Note~7). 

\section{Data availability}
The datasets generated and/or analyzed during the current study are not publicly available due to their forming part of an ongoing follow-up investigation but are available from the corresponding author on reasonable request.

\section{Code availability}
The code generated and/or analyzed during the current study are not publicly available due to its forming part of an ongoing follow-up investigation but are available from the corresponding author on reasonable request.

\section{Acknowledgements}
A.C., Z.-S.X., G.K., and A.W.E. acknowledge support from the Knut and Alice Wallenberg (KAW) Foundation through the Wallenberg Centre for Quantum Technology (WACQT). J.G. acknowledges support from Swedish Research Council (Ref: 2023-06671 and 2023-05288), Vinnova project (Ref: 2024-00466) and the Göran Gustafsson Foundation. A.W.E. acknowledges support from Swedish Research Council (VR) Starting Grant (Ref: 2016-03905). V.Z. acknowledges support from the KAW. 
I.M.K. acknowledges support by the French National Research Agency (ANR) under the France 2030 program, 
in subprograms "Choose France for Science", reference ``ANR-25-CFFS-0019'', and 
the TIRIS (Toulouse Initiative for Research's Impact on Society) Blockbuster Programme, reference ``ANR-22-EXES-0015'', the Occitanie region, the European Regional Development Fund (ERDF), and the European Union NextGenerationEU.
The funders had no role in the study.

\section{Author contributions}
A.C. performed the experiments, simulations, data analysis, and manuscript drafting. I.M.K. and A.W.E. conceived the idea. A.C., R.Y., and Z.-S.X. developed the experimental platform. G.K. contributed to discussions on the unitary decomposition algorithm and calibration procedure. I.M.K., J.G., and A.W.E. discussed the results and contributed to the manuscript. A.W.E. and J.G. supervised the project. V.Z. provided input on the manuscript.

\section{Competing interests}
The authors declare no competing financial or non-financial interests.

\bibliography{references}

@article{Halperin1982,
  title = {Quantized Hall conductance, current-carrying edge states, and the existence of extended states in a two-dimensional disordered potential},
  author = {Halperin, B. I.},
  journal = {Phys. Rev. B},
  volume = {25},
  issue = {4},
  pages = {2185--2190},
  numpages = {0},
  year = {1982},
  month = {Feb},
  publisher = {American Physical Society},
  doi = {10.1103/PhysRevB.25.2185}
}

@article{Thouless1982,
  title = {Quantized Hall Conductance in a Two-Dimensional Periodic Potential},
  author = {Thouless, D. J. and Kohmoto, M. and Nightingale, M. P. and den Nijs, M.},
  journal = {Phys. Rev. Lett.},
  volume = {49},
  issue = {6},
  pages = {405--408},
  numpages = {0},
  year = {1982},
  month = {Aug},
  publisher = {American Physical Society},
  doi = {10.1103/PhysRevLett.49.405}
}

@article{Onur2011,
   title={Artificial gauge field for photons in coupled cavity arrays},
   volume={84},
   ISSN={1094-1622},
   DOI={10.1103/physreva.84.043804},
   number={4},
   journal={Physical Review A},
   publisher={American Physical Society (APS)},
   author={Umucalılar, R. O. and Carusotto, I.},
   year={2011},
   month=oct }

@article{Fang2012_1,
  title = {Photonic Aharonov-Bohm Effect Based on Dynamic Modulation},
  author = {Fang, Kejie and Yu, Zongfu and Fan, Shanhui},
  journal = {Phys. Rev. Lett.},
  volume = {108},
  issue = {15},
  pages = {153901},
  numpages = {5},
  year = {2012},
  month = {Apr},
  publisher = {American Physical Society},
  doi = {10.1103/PhysRevLett.108.153901}
}

@article{Hafezi2013,
   title={Imaging topological edge states in silicon photonics},
   volume={7},
   ISSN={1749-4893},
   DOI={10.1038/nphoton.2013.274},
   number={12},
   journal={Nature Photonics},
   publisher={Springer Science and Business Media LLC},
   author={Hafezi, M. and Mittal, S. and Fan, J. and Migdall, A. and Taylor, J. M.},
   year={2013},
   month=oct, pages={1001–1005} }

@article{Haldane2008,
  title = {Possible Realization of Directional Optical Waveguides in Photonic Crystals with Broken Time-Reversal Symmetry},
  author = {Haldane, F. D. M. and Raghu, S.},
  journal = {Phys. Rev. Lett.},
  volume = {100},
  issue = {1},
  pages = {013904},
  numpages = {4},
  year = {2008},
  month = {Jan},
  publisher = {American Physical Society},
  doi = {10.1103/PhysRevLett.100.013904}
}

@article{Raghu2008,
  title = {Analogs of quantum-Hall-effect edge states in photonic crystals},
  author = {Raghu, S. and Haldane, F. D. M.},
  journal = {Phys. Rev. A},
  volume = {78},
  issue = {3},
  pages = {033834},
  numpages = {21},
  year = {2008},
  month = {Sep},
  publisher = {American Physical Society},
  doi = {10.1103/PhysRevA.78.033834}
}

@article{Wang2009,
  title = {Observation of unidirectional backscattering-immune topological electromagnetic states},
  author = {Wang, Zheng and Chong, Yidong and Joannopoulos, J. D. and Solja{\v{c}}i{\'c}, Marin},
  journal = {Nature},
  volume = {461},
  number = {7265},
  pages = {772--775},
  year = {2009},
  doi = {10.1038/nature08293}
}

@article{Rechtsman2013,
  author = {Rechtsman, Mikael C. and Zeuner, Julia M. and Tünnermann, Andreas and Nolte, Stefan and Segev, Mordechai and Szameit, Alexander},
  title = {Strain-induced pseudomagnetic field and photonic Landau levels in dielectric structures},
  journal = {Nature Photonics},
  volume = {7},
  number = {2},
  pages = {153--158},
  year = {2013},
  doi = {10.1038/nphoton.2012.302}
}

@article{Jamadi2020,
  author = {Jamadi, Omar and Rozas, Elena and Salerno, Grazia and Milićević, Marijana and Ozawa, Tomoki and Sagnes, Isabelle and Lemaître, Aristide and Le Gratiet, Luc and Harouri, Abdelmounaim and Carusotto, Iacopo and Bloch, Jacqueline and Amo, Alberto},
  title = {Direct observation of photonic Landau levels and helical edge states in strained honeycomb lattices},
  journal = {Light: Science \& Applications},
  volume = {9},
  number = {1},
  pages = {144},
  year = {2020},
  doi = {10.1038/s41377-020-00377-6}
}

@article{Guglielmon2021,
  title = {Landau levels in strained two-dimensional photonic crystals},
  author = {Guglielmon, J. and Rechtsman, M. C. and Weinstein, M. I.},
  journal = {Phys. Rev. A},
  volume = {103},
  issue = {1},
  pages = {013505},
  numpages = {16},
  year = {2021},
  month = {Jan},
  publisher = {American Physical Society},
  doi = {10.1103/PhysRevA.103.013505}
}

@article{Barczyk2024,
  author = {Barczyk, René and Kuipers, L. and Verhagen, Ewold},
  title = {Observation of Landau levels and chiral edge states in photonic crystals through pseudomagnetic fields induced by synthetic strain},
  journal = {Nature Photonics},
  volume = {18},
  number = {6},
  pages = {574--579},
  year = {2024},
  doi = {10.1038/s41566-024-01412-3}
}

@article{Barsukova2024,
  author = {Barsukova, Maria and Grisé, Fabien and Zhang, Zeyu and Vaidya, Sachin and Guglielmon, Jonathan and Weinstein, Michael I. and He, Li and Zhen, Bo and McEntaffer, Randall and Rechtsman, Mikael C.},
  title = {Direct observation of Landau levels in silicon photonic crystals},
  journal = {Nature Photonics},
  volume = {18},
  number = {6},
  pages = {580--585},
  year = {2024},
  doi = {10.1038/s41566-024-01425-y}
}

@article{Fang2012_2,
  author = {Fang, Kejie and Yu, Zongfu and Fan, Shanhui},
  title = {Realizing effective magnetic field for photons by controlling the phase of dynamic modulation},
  journal = {Nature Photonics},
  volume = {6},
  number = {11},
  pages = {782--787},
  year = {2012},
  month = {11},
  doi = {10.1038/nphoton.2012.236}
}

@article{Tzuang2014,
  author = {Tzuang, Lawrence D. and Fang, Kejie and Nussenzveig, Paulo and Fan, Shanhui and Lipson, Michal},
  title = {Non-reciprocal phase shift induced by an effective magnetic flux for light},
  journal = {Nature Photonics},
  volume = {8},
  number = {9},
  pages = {701--705},
  year = {2014},
  month = {9},
  doi = {10.1038/nphoton.2014.177}
}

@article{Schmidt2015,
author = {M. Schmidt and S. Kessler and V. Peano and O. Painter and F. Marquardt},
journal = {Optica},
number = {7},
pages = {635--641},
publisher = {Optica Publishing Group},
title = {Optomechanical creation of magnetic fields for photons on a lattice},
volume = {2},
month = {Jul},
year = {2015},
doi = {10.1364/OPTICA.2.000635}
}

@article{Dutt2020,
  author = {Avik Dutt  and Qian Lin  and Luqi Yuan  and Momchil Minkov  and Meng Xiao  and Shanhui Fan },
  title = {A single photonic cavity with two independent physical synthetic dimensions},
  journal = {Science},
  volume = {367},
  number = {6473},
  pages = {59-64},
  year = {2020},
  doi = {10.1126/science.aaz3071}
}

@article{Dinh2024,
  author = {Dinh, Hiep X. and Balčytis, Armandas and Ozawa, Tomoki and Ota, Yasutomo and Ren, Guanghui and Baba, Toshihiko and Iwamoto, Satoshi and Mitchell, Arnan and Nguyen, Thach G.},
  title = {Reconfigurable synthetic dimension frequency lattices in an integrated lithium niobate ring cavity},
  journal = {Communications Physics},
  volume = {7},
  number = {1},
  pages = {185},
  year = {2024},
  month = {6},
  day = {11},
  doi = {10.1038/s42005-024-01676-9}
}

@article{Reck94,
  title = {Experimental realization of any discrete unitary operator},
  author = {Reck, Michael and Zeilinger, Anton and Bernstein, Herbert J. and Bertani, Philip},
  journal = {Phys. Rev. Lett.},
  volume = {73},
  issue = {1},
  pages = {58--61},
  numpages = {0},
  year = {1994},
  month = {Jul},
  publisher = {American Physical Society},
  doi = {10.1103/PhysRevLett.73.58}
}

@article{Clements16,
author = {William R. Clements and Peter C. Humphreys and Benjamin J. Metcalf and W. Steven Kolthammer and Ian A. Walmsley},
journal = {Optica},
number = {12},
pages = {1460--1465},
publisher = {Optica Publishing Group},
title = {Optimal design for universal multiport interferometers},
volume = {3},
month = {Dec},
year = {2016},
doi = {10.1364/OPTICA.3.001460},
}

@article{Carolan15,
 author = {Carolan, Jacques and Harrold, Christopher and Sparrow, Chris and Mart{\'\i}n-López, Enrique and Russell, Nicholas J and Silverstone, Joshua W and Shadbolt, Peter J and Matsuda, Nobuyuki and Oguma, Manabu and Itoh, Mikitaka and others},
 journal = {Science},
 number = {6249},
 pages = {711--716},
 publisher = {American Association for the Advancement of Science},
 title = {Universal linear optics},
 volume = {349},
 year = {2015}
}

@article{Wang2020,
  author = {Wang, Jianwei and Sciarrino, Fabio and Laing, Anthony and Thompson, Mark G.},
  title = {Integrated photonic quantum technologies},
  journal = {Nature Photonics},
  volume = {14},
  number = {5},
  pages = {273--284},
  year = {2020},
  month = {5},
  issn = {1749-4893}
}

@article{Qiang2018,
  author = {Qiang, Xiaogang and Zhou, Xiaoqi and Wang, Jianwei and Wilkes, Callum M. and Loke, Thomas and O'Gara, Sean and Kling, Laurent and Marshall, Graham D. and Santagati, Raffaele and Ralph, Timothy C. and Wang, Jingbo B. and O'Brien, Jeremy L. and Thompson, Mark G. and Matthews, Jonathan C. F.},
  title = {Large-scale silicon quantum photonics implementing arbitrary two-qubit processing},
  journal = {Nature Photonics},
  volume = {12},
  number = {9},
  pages = {534--539},
  year = {2018},
  month = {9},
  issn = {1749-4893}
}

@article{Paesani2019,
  author = {Paesani, Stefano and Ding, Yunhong and Santagati, Raffaele and Chakhmakhchyan, Levon and Vigliar, Caterina and Rottwitt, Karsten and Oxenløwe, Leif K. and Wang, Jianwei and Thompson, Mark G. and Laing, Anthony},
  title = {Generation and sampling of quantum states of light in a silicon chip},
  journal = {Nature Physics},
  volume = {15},
  number = {9},
  pages = {925--929},
  year = {2019},
  month = {9},
  issn = {1745-2481}
}

@article{Shen2017,
  author = {Shen, Yichen and Harris, Nicholas C. and Skirlo, Scott and Prabhu, Mihika and Baehr-Jones, Tom and Hochberg, Michael and Sun, Xin and Zhao, Shijie and Larochelle, Hugo and Englund, Dirk and Soljačić, Marin},
  title = {Deep learning with coherent nanophotonic circuits},
  journal = {Nature Photonics},
  volume = {11},
  number = {7},
  pages = {441--446},
  year = {2017},
  month = {7},
  issn = {1749-4893}
}

@article{Shastri2021,
  author = {Shastri, Bhavin J. and Tait, Alexander N. and {Ferreira de Lima}, T. and Pernice, Wolfram H. P. and Bhaskaran, Harish and Wright, C. D. and Prucnal, Paul R.},
  title = {Photonics for artificial intelligence and neuromorphic computing},
  journal = {Nature Photonics},
  volume = {15},
  number = {2},
  pages = {102--114},
  year = {2021},
  month = {2},
  issn = {1749-4893}
}

@article{Marpaung2019,
  author = {Marpaung, David and Yao, Jianping and Capmany, José},
  title = {Integrated microwave photonics},
  journal = {Nature Photonics},
  volume = {13},
  number = {2},
  pages = {80--90},
  year = {2019},
  month = {2},
  issn = {1749-4893}
}

@article{Perez2017,
  author = {Pérez, Daniel and Gasulla, Ivana and Crudgington, Lee and Thomson, David J. and Khokhar, Ali Z. and Li, Ke and Cao, Wei and Mashanovich, Goran Z. and Capmany, José},
  title = {Multipurpose silicon photonics signal processor core},
  journal = {Nature Communications},
  volume = {8},
  number = {1},
  pages = {636},
  year = {2017},
  month = {9},
  day = {21},
  issn = {2041-1723}
}

@article{Bogaerts2020,
  author = {Bogaerts, Wim and Pérez, Daniel and Capmany, José and Miller, David A. B. and Poon, Joyce and Englund, Dirk and Morichetti, Francesco and Melloni, Andrea},
  title = {Programmable photonic circuits},
  journal = {Nature},
  volume = {586},
  number = {7828},
  pages = {207--216},
  year = {2020},
  month = {10},
  issn = {1476-4687}
}

@article{Harris2017,
  author = {Harris, Nicholas C. and Steinbrecher, Gregory R. and Prabhu, Mihika and Lahini, Yoav and Mower, Jacob and Bunandar, Darius and Chen, Changchen and Wong, Franco N. C. and Baehr-Jones, Tom and Hochberg, Michael and Lloyd, Seth and Englund, Dirk},
  title = {Quantum transport simulations in a programmable nanophotonic processor},
  journal = {Nature Photonics},
  volume = {11},
  number = {7},
  pages = {447--452},
  year = {2017},
  month = {7},
  issn = {1749-4893}
}

@article{On2024,
  author = {On, Mehmet Berkay and Ashtiani, Farshid and Sanchez-Jacome, David and Perez-Lopez, Daniel and Yoo, S. J. Ben and Blanco-Redondo, Andrea},
  title = {Programmable integrated photonics for topological Hamiltonians},
  journal = {Nature Communications},
  year = {2024},
  month = {Jan},
  day = {20},
  volume = {15},
  number = {1},
  pages = {629},
  doi = {10.1038/s41467-024-44939-3},
  issn = {2041-1723}
}

@article{Dai2024,
  author = {Dai, Tianxiang and Ma, Anqi and Mao, Jun and Ao, Yutian and Jia, Xinyu and Zheng, Yun and Zhai, Chonghao and Yang, Yan and Li, Zhihua and Tang, Bo and Luo, Jun and Zhang, Baile and Hu, Xiaoyong and Gong, Qihuang and Wang, Jianwei},
  title = {A programmable topological photonic chip},
  journal = {Nature Materials},
  year = {2024},
  month = {Jul},
  volume = {23},
  number = {7},
  pages = {928--936},
  doi = {10.1038/s41563-024-01904-1},
  issn = {1476-4660}
}

@article{Ma2024,
author = {Anqi Ma and Tianxiang Dai and Jun Mao and Zhaorong Fu and Yan Yang and Xiaoyong Hu and Qihuang Gong and Jianwei Wang},
journal = {Optica},
number = {11},
pages = {1533--1539},
publisher = {Optica Publishing Group},
title = {Anisotropic quantum transport in a programmable photonic topological insulator},
volume = {11},
month = {Nov},
year = {2024},
doi = {10.1364/OPTICA.539301}
}

@article{Love2025,
title = {A programmable platform for photonic topological insulators},
title = {},
author = {Stuart Love and Mohamad Hossein Idjadi and Farshid Ashtiani and Howard (Ho-Wai) Lee and Andrea Blanco-Redondo},
pages = {367--373},
volume = {14},
number = {3},
journal = {Nanophotonics},
doi = {doi:10.1515/nanoph-2024-0577},
year = {2025},
lastchecked = {2025-12-22}
}

@article{Sparrow2018,
  author = {Sparrow, Chris and others},
  title = {Simulating the vibrational quantum dynamics of molecules using photonics},
  journal = {Nature},
  volume = {557},
  pages = {660--667},
  year = {2018}
}

@article{Qiang2021,
  author = {Qiang, Xiaogang and others},
  title = {Implementing graph-theoretic quantum algorithms on a silicon photonic quantum walk processor},
  journal = {Science Advances},
  volume = {7},
  pages = {eabb8375},
  year = {2021}
}

@article{Wang2022,
  author = {Wang, Yang and Yu, Xinyao and Xue, Shichuan and Wang, Yizhi and Zhan, Junwei and Wu, Chao and Zhu, Pingyu and Zheng, Qilin and Yu, Miaomiao and Liu, Yingwen and Qiang, Xiaogang and Wu, Junjie and Yang, Xuejun and Xu, Ping},
  title = {Experimental demonstration of quantum transport enhancement using time-reversal symmetry breaking on a silicon photonic chip},
  journal = {Science China Physics, Mechanics \& Astronomy},
  volume = {65},
  pages = {100362},
  year = {2022}
}

@article{Maring2024,
  author = {Maring, Nicolas and others},
  title = {A versatile single-photon-based quantum computing platform},
  journal = {Nature Photonics},
  volume = {18},
  pages = {603--609},
  year = {2024}
}

@article{Taballione2023,
  author = {Taballione, Caterina and others},
  title = {20-mode universal quantum photonic processor},
  journal = {Quantum},
  volume = {7},
  pages = {1071},
  year = {2023}
}

@article{Rechtsman2013Floquet,
  author = {Rechtsman, Mikael C. and Zeuner, Julia M. and Plotnik, Yonatan and Lumer, Yaakov and Podolsky, Daniel and Dreisow, Felix and Nolte, Stefan and Segev, Mordechai and Szameit, Alexander},
  title = {Photonic {Floquet} topological insulators},
  journal = {Nature},
  volume = {496},
  pages = {196--200},
  year = {2013}
}

@article{Maczewsky2017,
  author = {Maczewsky, Lukas J. and Zeuner, Julia M. and Nolte, Stefan and Szameit, Alexander},
  title = {Observation of photonic anomalous {Floquet} topological insulators},
  journal = {Nature Communications},
  volume = {8},
  pages = {13756},
  year = {2017}
}

@article{Aaronson2013,
  author = {Aaronson, Scott and Arkhipov, Alex},
  title = {The computational complexity of linear optics},
  journal = {Theory of Computing},
  volume = {9},
  pages = {143--252},
  year = {2013}
}

@article{Tischler2018,
  author = {Tischler, Nora and Rockstuhl, Carsten and S{\l}owik, Karolina},
  title = {Quantum optical realization of arbitrary linear transformations allowing for loss and gain},
  journal = {Physical Review X},
  volume = {8},
  pages = {021017},
  year = {2018}
}

@article{Bandyopadhyay2024,
  author = {Bandyopadhyay, Saumil and others},
  title = {Single-chip photonic deep neural network with forward-only training},
  journal = {Nature Photonics},
  volume = {18},
  pages = {1335--1343},
  year = {2024}
}

@article{Taballione2021,
  author = {Taballione, Caterina and others},
  title = {A universal fully reconfigurable 12-mode quantum photonic processor},
  journal = {Materials for Quantum Technology},
  volume = {1},
  pages = {035002},
  year = {2021}
}

@article{Krishna2026,
  author = {Krishna, Govind and Yadgirkar, Rohan and Krishnakumar, Balakrishnan and Cataldo, Andrea and Xu, Ze-Sheng and Los, Johannes W. N. and Zwiller, Val and Elshaari, Ali W. and Gao, Jun},
  title = {Experimental simulation of a non-equilibrium quantum piston on a programmable photonic quantum computer},
  journal = {Optica},
  volume = {13},
  pages = {2105--2117},
  year = {2026}
}

@article{Cataldo2026,
  title = {Interferometric Readout of Momentum-Space Topology in a Programmable Dissipative Photonic Circuit},
  author = {Cataldo, Andrea and Bergholtz, Emil J. and Leykam, Daniel and Gao, Jun and Elshaari, Ali W.},
  journal = {Phys. Rev. Lett.},
  volume = {137},
  issue = {20},
  pages = {206601},
  numpages = {6},
  year = {2026},
  month = {Nov},
  publisher = {American Physical Society},
  doi = {10.1103/xkht-64kc}
}

\clearpage
\section{\figuresectiontitle}

\mainfigure{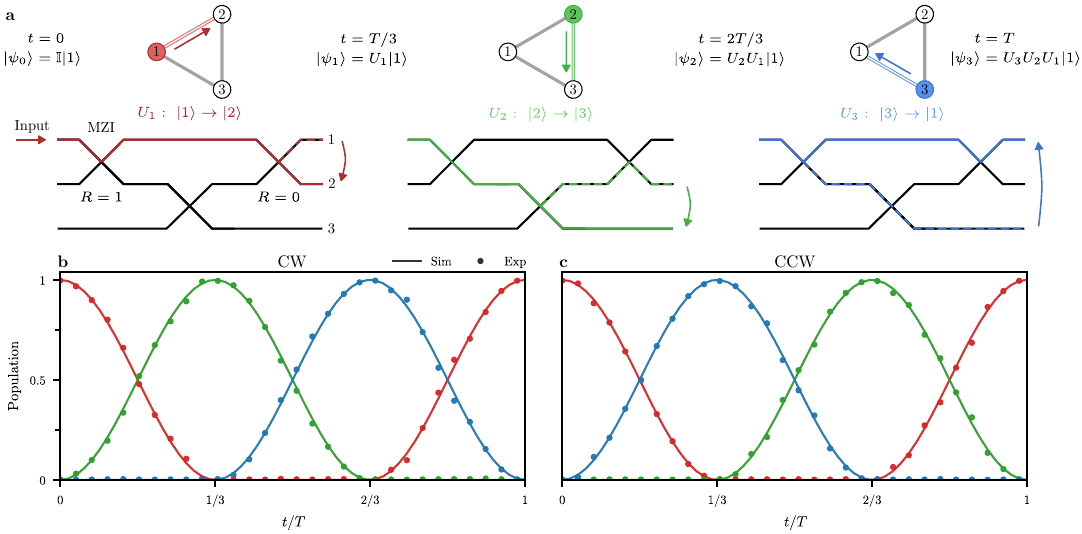}
\captionof{figure}{\textbf{Chiral Floquet micromotion in a triangular plaquette.} \textbf{a)} Programmed Rabi-transfer sequence for CW modulation; colored bonds indicate the active substep, while time labels denote interval boundaries. The plaquette sites correspond to spatial waveguide modes, with crossings denoting tunable MZIs that set the effective coupling between the modes. The reflectivity \(R\) of an MZI is tuned between full reflection \(R=1\) and full transfer \(R=0\), realizing a Rabi \(\pi\)-pulse; \(U_1(T/3)\) transfers \(1\to2\), \(U_2(T/3)\) transfers \(2\to3\), and \(U_3(T/3)\) transfers \(3\to1\). \textbf{b)} CW modulation; the population follows \(1\to2\to3\to1\), with target-site populations \(P_1(0)=99.65\pm0.07\%\), \(P_2(T/3)=99.51\pm0.04\%\), \(P_3(2T/3)=99.63\pm0.05\%\), and \(P_1(T)=99.15\pm0.03\%\). \textbf{c)} CCW modulation; reversing the drive order reverses the circulation to \(1\to3\to2\to1\), with target-site populations \(P_1(0)=99.63\pm0.02\%\), \(P_3(T/3)=99.42\pm0.04\%\), \(P_2(2T/3)=99.05\pm0.15\%\), and \(P_1(T)=99.25\pm0.09\%\). Solid lines denote simulations; markers denote experiments.}
\label{fig:Figure_1}

\mainfigure{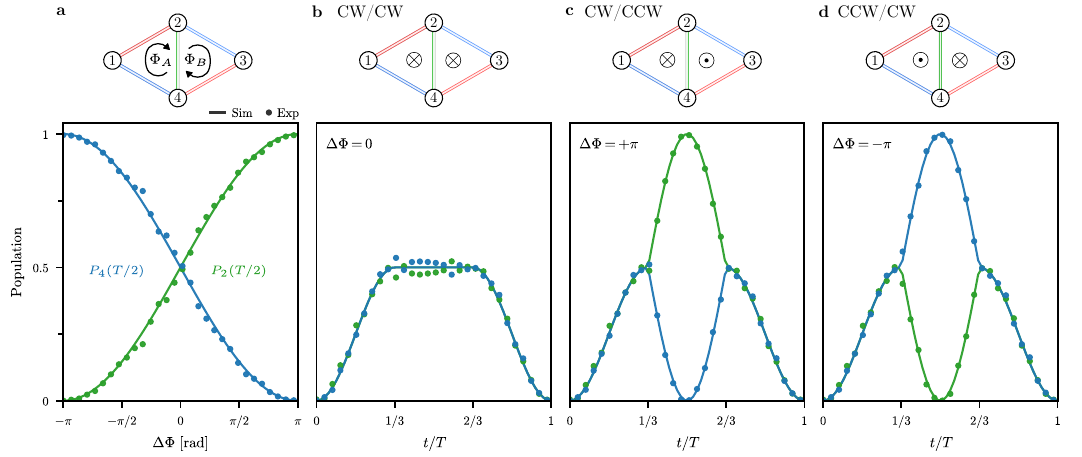}
\captionof{figure}{\textbf{Flux-tuned interference in a four-site lattice.} \textbf{a)} Schematic showing the two loop phases \(\Phi_A\) and \(\Phi_B\); for the input \((|1\rangle +|3\rangle)/\sqrt{2}\), the synthetic flux difference \(\Delta \Phi \equiv \Phi_A - \Phi_B\) steers interference at the shared sites 2 and 4. The mid-period populations \(P_2(T/2)\) (green) and \(P_4(T/2)\) (blue) are complementary as \(\Delta \Phi\) is varied. \textbf{b)} \(\Delta \Phi=0\), CW/CW; the population is split nearly equally, with \(P_2(T/2)=48.10 \pm 0.18 \%\) and \(P_4(T/2)=51.68 \pm 0.16 \%\). \textbf{c)} \(\Delta \Phi=\pi\), CW/CCW; interference routes the population almost entirely to site 2, with \(P_2(T/2)=99.59 \pm 0.05 \%\) and \(P_4(T/2)=0.16 \pm 0.04 \%\). \textbf{d)} \(\Delta \Phi=-\pi\), CCW/CW; the routing is reversed, with \(P_2(T/2)=0.046 \pm 0.002 \%\) and \(P_4(T/2)=99.73 \pm 0.03 \%\). In the latter two cases, destructive interference suppresses population transfer through the green coupling, so the effective dynamics proceed primarily through the red and blue couplings. Solid lines denote simulations; markers denote experiments.}
\label{fig:Figure_2}

\mainfigure{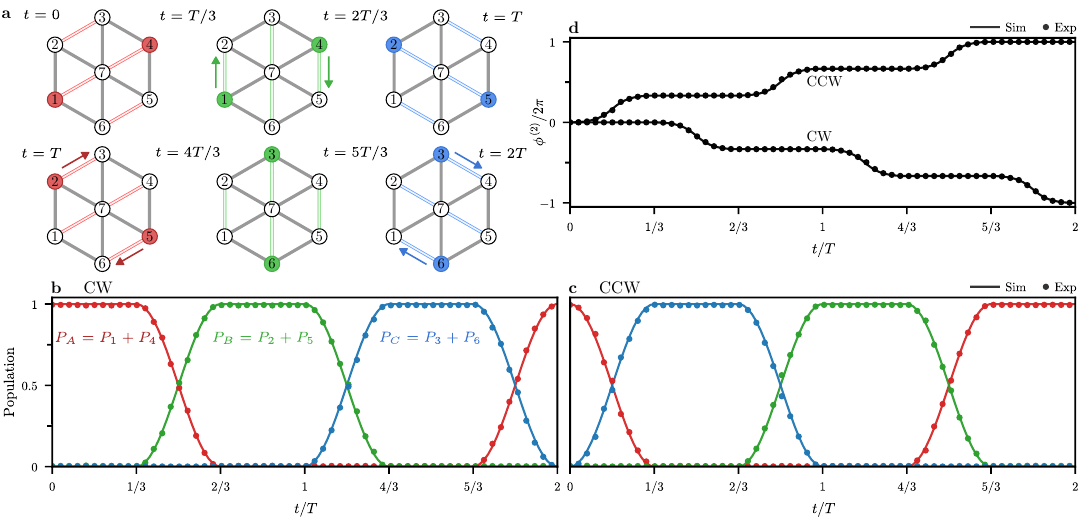}
\captionof{figure}{\textbf{Rabi-driven circulation in a seven-site hexagon.} \textbf{a)} Programmed sequence under CW modulation at the Rabi \(\pi\)-pulse condition; colored bonds indicate the active substep, while time labels denote interval boundaries. Opposite perimeter sites are grouped into sectors \(A=(1,4)\), \(B=(2,5)\), and \(C=(3,6)\); the system is initialized in the antisymmetric state \((|1\rangle-|4\rangle)/\sqrt{2}\). \textbf{b)} CW modulation; the grouped populations \(P_A=P_1+P_4\), \(P_B=P_2+P_5\), and \(P_C=P_3+P_6\) follow the sequence \(A \to B \to C \to A\) up to two modulation periods \(2T\), with measured target populations \(P_A(0)=99.58 \pm 0.05\%\), \(P_B(2T/3)=99.33 \pm 0.07\%\), \(P_C(4T/3)=99.21 \pm 0.02\%\), and \(P_A(2T)=99.54 \pm 0.07\%\). \textbf{c)} CCW modulation; reversing the drive order reverses the circulation, giving measured target populations \(P_A(0)=99.48\pm0.13\%\), \(P_C(T/3)=99.60\pm0.02\%\), \(P_B(T)=99.52\pm0.04\%\), and \(P_A(5T/3)=99.36\pm0.08\%\). The RMSE between experiment and simulation is \(7.61 \times 10^{-3}\) for CW and \(7.55 \times 10^{-3}\) for CCW. \textbf{d)} Second-harmonic phase \(\phi^{(2)}/2\pi\), computed from the six perimeter sites using Eq.~\eqref{eq:phi_2}. The phase drifts with opposite slopes for CW and CCW modulations. Linear fits give \(\nu_2^{\mathrm{CW}}=-0.504\) and \(\nu_2^{\mathrm{CCW}}=+0.506\). Solid lines denote simulations; markers denote experiments.}
\label{fig:Figure_3}

\mainfigure{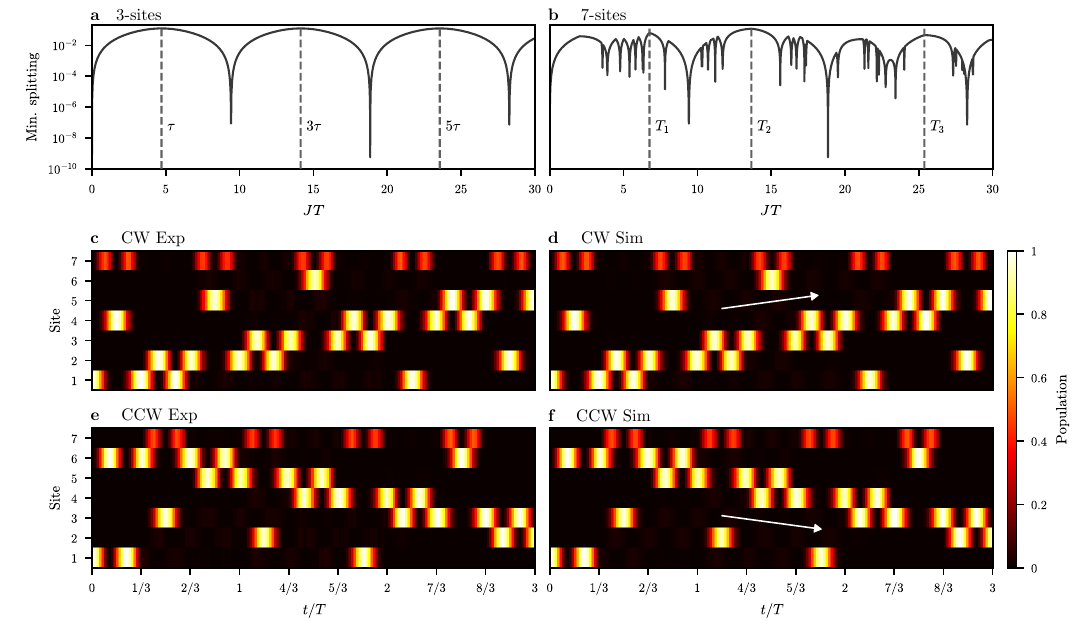}
\captionof{figure}{\textbf{Floquet-period optimization and single-site transport in a seven-site hexagon.} The system is initialized in state \(|1\rangle\). \textbf{a,b)} Minimal quasi-energy splitting \(\Delta \varepsilon_\text{min}\) of the Floquet operator \(U_F(T)\) versus modulation period \(JT\). Peaks at \(\tau\), \(3\tau\), \(5\tau\), with \(\tau=3\pi/(2J)\), mark the Rabi \(\pi\)-pulse condition and its odd multiples with the peak splitting reaching \(\Delta \varepsilon_\text{min} = 0.125\). The dips indicate near-degeneracy of quasi-energies. In the seven-site case, operating points are identified at \(T_1=6.7574/J\), \(T_2=13.6647/J\) (used in sim/exp, \(\Delta \varepsilon_\text{min} = 0.119\)), and \(T_3=25.3791/J\). \textbf{c,d)} CW modulation; experimental and simulated population dynamics over three driving periods (\(t=3T\)). Oscillatory exchanges are observed within each substep with an overall CW circulation (upward drift). \textbf{e,f)} CCW modulation; the circulation reverses (downward drift). The RMSE between experiment and simulation is \(8.65 \times 10^{-3}\) for CW and \(8.01 \times 10^{-3}\) for CCW. The second-harmonic analysis gives \(\nu_2^{\text{CW}}=-0.495\) and \(\nu_2^{\text{CCW}}=0.497\).}
\label{fig:Figure_4}

\mainfigure{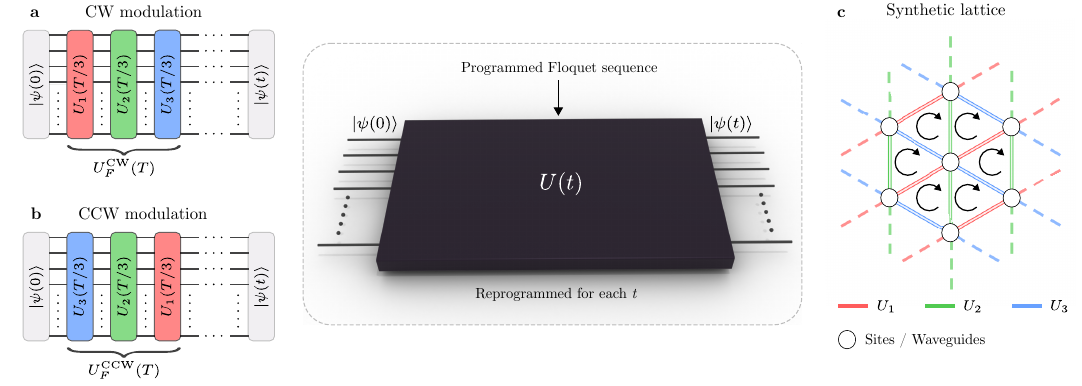}
\captionof{figure}{\textbf{Discretized Floquet evolution on a programmable photonic processor.} A state \(|\psi(0)\rangle\) injected on the left bus enters the reconfigurable MZI mesh, which implements the time-evolution operator \(U(t)\) on \(N\) spatial waveguide modes, yielding \(|\psi(t)\rangle=U(t)|\psi(0)\rangle\) at the right bus. \textbf{a)} CW modulation; one period \(T\) consists of three ordered substeps \(U_1(T/3)\) (red), \(U_2(T/3)\) (green), and \(U_3(T/3)\) (blue). The colored blocks show this programmed sequence of unitaries, not separate physical sections of the mesh. Their time-ordered product gives the Floquet operator \(U_F^{\text{CW}}(T)=U_3U_2U_1\), with horizontal dots indicating repetition over successive periods. Intermediate sampled times correspond to partial progress through the sequence; for example, \(t=T/3\) contains only \(U_1(T/3)\), whereas \(t=2T/3\) contains \(U_2(T/3)U_1(T/3)\). \textbf{b)} CCW modulation; obtained by reversing the substeps, giving \(U_F^{\text{CCW}}(T)=U_1U_2U_3\). \textbf{c)} Scalable \(N\)-site synthetic lattice; circles denote spatial waveguide modes, solid colored bonds indicate the couplings activated during each substep, and dashed extensions show how the pattern can be tiled to larger lattices.}
\label{fig:Figure_5}

\mainfigure{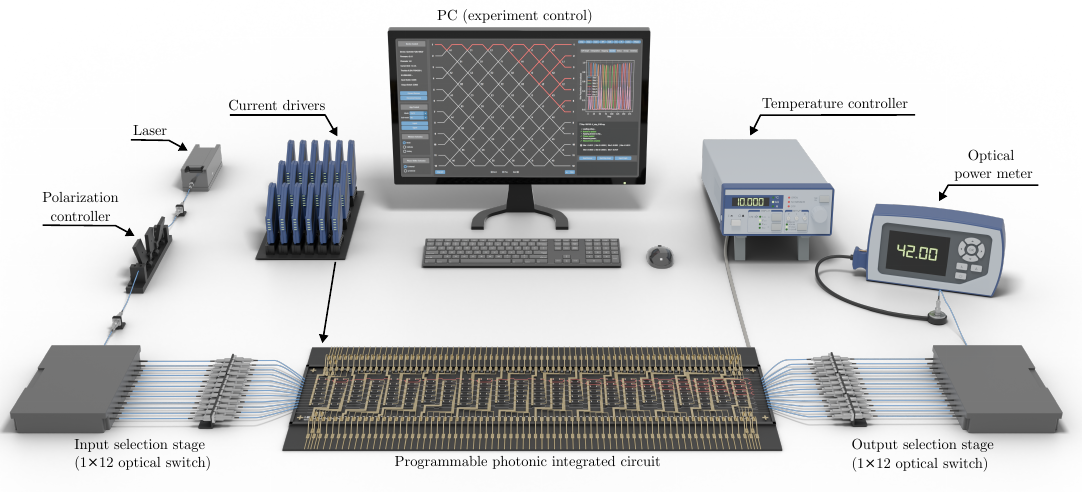}
\captionof{figure}{\textbf{Experimental setup.} Coherent light at 1550 nm from a continuous-wave laser passes through a polarization controller set for TE polarization and is directed to the common port of a \(1\times12\) MEMS optical switch. The switch can be programmed to route the light to any of its twelve multiplexed ports. These ports couple via edge couplers to the corresponding input waveguides of a 12-mode programmable photonic chip, thereby selecting the input. The chip consists of a reconfigurable MZI mesh driven by multichannel current drivers. Light at the chip output is coupled via edge couplers into a second switch operated in reverse; its common port is connected to an optical power meter. Control and acquisition are handled by a PC, which commands the switches, sets the heater currents to realize the unitaries, and logs readings from the power meter. A thermoelectric cooler beneath the chip, regulated by a PID temperature controller, maintains a fixed room-temperature setpoint.}
\label{fig:Figure_6}

\end{document}